\documentclass[twocolumn, tighten]{aastex63}
\usepackage{hyperref}
\usepackage{textcomp}
\usepackage{bm,color}
\usepackage{amsmath}
\usepackage{ulem}
\usepackage{csquotes}
\usepackage[T1]{fontenc}
\newcommand{\pb}{\textsc{Polarbear}}

\newcommand{\sptpol}{{\sc SPTpol}}

\newcommand{\actpol}{{\sc ACTpol}}

\newcommand{\planckeight}{{\it Planck}~2018}
\newcommand{\planck}{{\it Planck}}
\newcommand{\apex}{{\sc apex}}

\newcommand{\quiet}{{\sc quiet}}
\newcommand{\bicep}{{BICEP}}

\newcommand{\lcdm}{$\Lambda$CDM}
\newcommand{\keckarray}{\textit{Keck Array}}

\newcommand{\bmodes}{$B$-modes}
\newcommand{\bmode}{$B$-mode}
\newcommand{\emodes}{$E$-modes}
\newcommand{\emode}{$E$-mode}
\newcommand{\sqdeg}{\ensuremath{{\rm deg}^2}}
\newcommand{\comment}[1]{}
\newcommand{\tbd}[1]{{\color{red}{#1}}}
 \newcommand{\ho}{\ensuremath{H_0}}
 \newcommand{\uksq}{\ensuremath{\mu {\rm K}^2}}
 \newcommand{\neff}{\ensuremath{N_{\rm eff}}}
  \newcommand{\yhe}{\ensuremath{Y_{\rm He}}}

\newcommand{\mapdepth}{$32\,\mu\mathrm{K}$-$\mathrm{arcmin}$}

\newcommand{\patchloc}{(RA, Dec)=($+0^\mathrm{h}12^\mathrm{m}0^\mathrm{s},-59^\circ18^\prime$)}

\newcommand{\lcdmPlanckombh}{\ensuremath{  0.02237 \pm  0.00015  }}
\newcommand{\lcdmPlanckomch}{\ensuremath{   0.1200 \pm   0.0014  }}
\newcommand{\lcdmPlancktheta}{\ensuremath{  1.04089 \pm  0.00032  }}
\newcommand{\lcdmPlancktau}{\ensuremath{   0.0548 \pm   0.0079  }}
\newcommand{\lcdmPlanckns}{\ensuremath{   0.9650 \pm   0.0045  }}
\newcommand{\lcdmPlancklogA}{\ensuremath{    3.044 \pm    0.016  }}
\newcommand{\lcdmPlanckho}{\ensuremath{    67.42 \pm     0.69  }}

\newcommand{\yhePlanckyhe}{\ensuremath{    0.239 \pm    0.013  }}

\newcommand{\lcdmCMBombh}{\ensuremath{  0.02230 \pm  0.00014  }}
\newcommand{\lcdmCMBomch}{\ensuremath{   0.1203 \pm   0.0013  }}
\newcommand{\lcdmCMBtheta}{\ensuremath{  1.04102 \pm  0.00030  }}
\newcommand{\lcdmCMBtau}{\ensuremath{   0.0527 \pm   0.0070  }}
\newcommand{\lcdmCMBns}{\ensuremath{   0.9636 \pm   0.0042  }}
\newcommand{\lcdmCMBlogA}{\ensuremath{    3.043 \pm    0.015  }}
\newcommand{\lcdmCMBho}{\ensuremath{    67.20 \pm     0.57  }}
\newcommand{\lcdmCMBypbbn}{\ensuremath{ 0.246696 \pm 0.000062  }}

\newcommand{\neffCMBnnu}{\ensuremath{     2.94 \pm     0.16  }}

\newcommand{\yheCMByhe}{\ensuremath{    0.248 \pm    0.012  }}

\newcommand{\yheneffCMByhe}{\ensuremath{    0.262 \pm    0.015  }}
\newcommand{\yheneffCMBnnu}{\ensuremath{     2.70 \pm     0.26  }}

\newcommand{\lcdmCMBextombh}{\ensuremath{  0.02234 \pm  0.00015  }}
\newcommand{\lcdmCMBextomch}{\ensuremath{   0.1198 \pm   0.0011  }}
\newcommand{\lcdmCMBexttheta}{\ensuremath{  1.04104 \pm  0.00030  }}
\newcommand{\lcdmCMBexttau}{\ensuremath{   0.0518 \pm   0.0065  }}
\newcommand{\lcdmCMBextns}{\ensuremath{   0.9647 \pm   0.0038  }}
\newcommand{\lcdmCMBextlogA}{\ensuremath{    3.040 \pm    0.014  }}
\newcommand{\lcdmCMBextho}{\ensuremath{    67.41 \pm     0.51  }}

\newcommand{\neffCMBextombh}{\ensuremath{  0.02220 \pm  0.00022  }}
\newcommand{\neffCMBextomch}{\ensuremath{   0.1177 \pm   0.0027  }}
\newcommand{\neffCMBexttheta}{\ensuremath{  1.04125 \pm  0.00040  }}
\newcommand{\neffCMBexttau}{\ensuremath{   0.0511 \pm   0.0075  }}
\newcommand{\neffCMBextns}{\ensuremath{   0.9592 \pm   0.0075  }}
\newcommand{\neffCMBextlogA}{\ensuremath{    3.031 \pm    0.016  }}
\newcommand{\neffCMBextho}{\ensuremath{     66.5 \pm      1.4  }}

\newcommand{\neffCMBextnnu}{\ensuremath{     2.90 \pm     0.18  }}
\newcommand{\yheCMBextombh}{\ensuremath{  0.02236 \pm  0.00019  }}
\newcommand{\yheCMBextomch}{\ensuremath{   0.1197 \pm   0.0011  }}
\newcommand{\yheCMBexttheta}{\ensuremath{  1.04104 \pm  0.00050  }}
\newcommand{\yheCMBexttau}{\ensuremath{   0.0511 \pm   0.0074  }}
\newcommand{\yheCMBextns}{\ensuremath{   0.9648 \pm   0.0066  }}
\newcommand{\yheCMBextlogA}{\ensuremath{    3.039 \pm    0.015  }}
\newcommand{\yheCMBextho}{\ensuremath{    67.49 \pm     0.59  }}

\newcommand{\yheCMBextyhe}{\ensuremath{    0.246 \pm    0.011  }}
\newcommand{\yheneffCMBextombh}{\ensuremath{  0.02221 \pm  0.00020  }}
\newcommand{\yheneffCMBextomch}{\ensuremath{   0.1138 \pm   0.0040  }}
\newcommand{\yheneffCMBexttheta}{\ensuremath{   1.0424 \pm   0.0011  }}
\newcommand{\yheneffCMBexttau}{\ensuremath{   0.0509 \pm   0.0073  }}
\newcommand{\yheneffCMBextns}{\ensuremath{   0.9580 \pm   0.0078  }}
\newcommand{\yheneffCMBextlogA}{\ensuremath{    3.028 \pm    0.016  }}
\newcommand{\yheneffCMBextho}{\ensuremath{     65.1 \pm      1.6  }}

\newcommand{\yheneffCMBextyhe}{\ensuremath{    0.263 \pm    0.014  }}
\newcommand{\yheneffCMBextnnu}{\ensuremath{     2.65 \pm     0.26  }}

\begin{document}

\title{A measurement of the CMB \emode{} angular power spectrum at subdegree scales from 670 square degrees of \pb{} data}

\correspondingauthor{Anh T.P. Pham}
\email{tpham5@student.unimelb.edu.au}

\author[0000-0002-0400-7555]{S. Adachi}
\affiliation{Department of Physics, Kyoto University, Kyoto 606-8502, Japan}

\author[0000-0002-1571-663X]{M. A. O. Aguilar Fa\'undez}
\affiliation{Department of Physics and Astronomy, Johns Hopkins University, Baltimore, MD 21218, USA}
\affiliation{Departamento de F\'isica, FCFM, Universidad de Chile, Blanco Encalada 2008, Santiago, Chile}

\author[0000-0002-3407-5305]{K. Arnold}
\affiliation{Department of Physics, University of California, San Diego, CA 92093-0424, USA}

\author[0000-0002-8211-1630]{C. Baccigalupi}
\affiliation{International School for Advanced Studies (SISSA), Via Bonomea 265, 34136, Trieste, Italy}
\affiliation{Institute for Fundamental Physics of the Universe (IFPU), Via Beirut 2, 34151, Grignano (TS), Italy}
\affiliation{The National Institute for Nuclear Physics, INFN, Sezione di Trieste Via Valerio 2, I-34127, Trieste, Italy}

\author[0000-0002-1623-5651]{D. Barron}
\affiliation{Department of Physics and Astronomy, University of New Mexico, Albuquerque, NM 87131, USA}

\author[0000-0003-0848-2756]{D. Beck}
\affiliation{Department of Physics, Stanford University, Stanford, CA 94305, USA}

%\author{S. Beckman}
%\affiliation{Department of Physics, University of California, Berkeley, CA 94720, USA}

\author[0000-0003-4847-3483]{F. Bianchini}
\affiliation{School of Physics, University of Melbourne, Parkville, VIC 3010, Australia}

%\author{D. Boettger}
%\affiliation{Instituto de Astrof\'isica and Centro de Astro-Ingenier\'ia, Facultad de F\'isica, Pontificia Universidad Cat\'olica de Chile, Av. Vicuna Mackenna 4860, 7820436 Macul, Santiago, Chile}

%\author{J. Borrill}
%\affiliation{Computational Cosmology Center, Lawrence Berkeley National Laboratory, Berkeley, CA 94720, USA}
%\affiliation{Space Sciences Laboratory, University of California, Berkeley, CA 94720, USA}

%\author[0000-0002-5751-1392]{J. Carron}
%\affiliation{Department of Physics \& Astronomy, University of Sussex, Brighton BN1 9QH, UK}

\author{S. Chapman}
\affiliation{Department of Physics and Atmospheric Science, Dalhousie University, Halifax, NS, B3H 4R2, Canada}

\author[0000-0002-7764-378X]{K. Cheung}
\affiliation{Department of Physics, University of California, Berkeley, CA 94720, USA}

\author[0000-0002-3266-857X]{Y. Chinone}

\affiliation{Research Center for the Early Universe, School of Science, The University of Tokyo, Tokyo 113-0033, Japan}
\affiliation{Kavli Institute for the Physics and Mathematics of the Universe (Kavli IPMU, WPI), UTIAS, The University of Tokyo, Kashiwa, Chiba 277-8583, Japan}
%\affiliation{Kavli Institute for the Physics and Mathematics of the Universe (Kavli IPMU, WPI), Berkeley Satellite, the University of California, Berkeley 94720, USA}

\author[0000-0001-5068-1295]{K. Crowley}
\affiliation{Department of Physics, University of California, Berkeley, CA 94720, USA}

%\author[0000-0002-7471-719X]{A. Cukierman}
%\affiliation{Kavli Institute for Particle Astrophysics and Cosmology,SLAC National Accelerator Laboratory,2575 Sand Hill Rd, Menlo Park , CA 94025}
%\affiliation{Department of Physics, Stanford University, Stanford, CA, 94305}

\author{M. Dobbs}
\affiliation{Physics Department, McGill University, Montreal, QC H3A 0G4, Canada}
\affiliation{Canadian Institute for Advance Research (CIfAR), Toronto, Canada, M5G 1M1}

\author[0000-0001-5471-3434]{H. El Bouhargani}
\affiliation{Universit\'e de Paris, CNRS, Astroparticule et Cosmologie, F-75013 Paris, France}

\author[0000-0002-5166-5614]{T. Elleflot}
\affiliation{Physics Division, Lawrence Berkeley National Laboratory, Berkeley, CA 94720, USA}

\author[0000-0002-1419-0031]{J. Errard}
\affiliation{Universit\'e de Paris, CNRS, Astroparticule et Cosmologie, F-75013 Paris, France}

\author[0000-0002-3255-4695]{G. Fabbian}
\affiliation{Department of Physics \& Astronomy, University of Sussex, Brighton BN1 9QH, UK}

\author{C. Feng}
\affiliation{Department of Physics, University of Illinois at Urbana-Champaign, 1110 W Green St, Urbana, IL, 61801, USA}

\author{T. Fujino}
\affiliation{Yokohama National University, Yokohama, Kanagawa 240-8501, Japan}

\author{N. Galitzki}
\affiliation{Department of Physics, University of California, San Diego, CA 92093-0424, USA}

\author{N. Goeckner-Wald}
\affiliation{Department of Physics, University of California, Berkeley, CA 94720, USA}
\affiliation{Department of Physics, Stanford University, Stanford, CA 94305, USA}

\author{J. Groh}
\affiliation{Department of Physics, University of California, Berkeley, CA 94720, USA}

\author{G. Hall}
\affiliation{Minnesota Institute for Astrophysics, University of Minnesota, Minneapolis, MN 55455, USA}

%\author{N. Halverson}
%\affiliation{Center for Astrophysics and Space Astronomy, University of Colorado, Boulder, CO 80309, USA}
%\affiliation{Department of Astrophysical and Planetary Sciences, University of Colorado, Boulder, CO 80309, USA}
%\affiliation{Department of Physics, University of Colorado, Boulder, CO 80309, USA}

%\author{T. Hamada}
%\affiliation{Astronomical Institute, Tohoku University, Sendai, Miyagi 980-0845, Japan}

\author[0000-0003-1443-1082]{M. Hasegawa}
\affiliation{Institute of Particle and Nuclear Studies (IPNS), High Energy Accelerator Research Organization (KEK), Tsukuba, Ibaraki 305-0801, Japan}

\author[0000-0001-6830-8309]{M. Hazumi}
\affiliation{Institute of Particle and Nuclear Studies (IPNS), High Energy Accelerator Research Organization (KEK), Tsukuba, Ibaraki 305-0801, Japan}
\affiliation{Kavli Institute for the Physics and Mathematics of the Universe (Kavli IPMU, WPI), UTIAS, The University of Tokyo, Kashiwa, Chiba 277-8583, Japan}
\affiliation{Institute of Space and Astronautical Science (ISAS), Japan Aerospace Exploration Agency (JAXA), Sagamihara, Kanagawa 252-0222, Japan}
\affiliation{SOKENDAI (The Graduate University for Advanced Studies), Shonan Village, Hayama, Kanagawa 240-0193, Japan}

%\author{C. A. Hill}
%\affiliation{Department of Physics, University of California, Berkeley, CA 94720, USA}
%\affiliation{Physics Division, Lawrence Berkeley National Laboratory, Berkeley, CA 94720, USA}
\author{H. Hirose}	
 \affiliation{Graduate School of Engineering Science, Yokohama National University, Yokohama 240-8501, Japan}
%\author{L. Howe}
%\affiliation{Department of Physics, University of California, San Diego, CA 92093-0424, USA}

%\author{Y. Inoue}
%\affiliation{Department of Physics, National Central University, Taoyuan 32002, Taiwan}
%\affiliation{Center for High Energy and High Field Physics, National Central University, Taoyuan 32002, Taiwan}
%\affiliation{High Energy Accelerator Research Organization (KEK), Tsukuba, Ibaraki 305-0801, Japan}

%\author[0000-0001-8697-0064]{G. Jaehnig}
%\affiliation{Center for Astrophysics and Space Astronomy, University of Colorado, Boulder, CO 80309, USA}
%\affiliation{Department of Astrophysical and Planetary Sciences, University of Colorado, Boulder, CO 80309, USA}
\author[0000-0003-2086-1759]{A. H. Jaffe}
\affiliation{Department of Physics, Imperial College London, London SW7 2AZ, United Kingdom}

\author[0000-0001-5893-7697]{O. Jeong}
\affiliation{Department of Physics, University of California, Berkeley, CA 94720, USA}

\author{D. Kaneko}
\affiliation{Kavli Institute for the Physics and Mathematics of the Universe (Kavli IPMU, WPI), UTIAS, The University of Tokyo, Kashiwa, Chiba 277-8583, Japan}

\author{N. Katayama}
\affiliation{Kavli Institute for the Physics and Mathematics of the Universe (Kavli IPMU, WPI), UTIAS, The University of Tokyo, Kashiwa, Chiba 277-8583, Japan}

\author[0000-0003-3118-5514]{B. Keating}
\affiliation{Department of Physics, University of California, San Diego, CA 92093-0424, USA}

%\author{R. Keskitalo}
%\affiliation{Computational Cosmology Center, Lawrence Berkeley National Laboratory, Berkeley, CA 94720, USA}
%\affiliation{Department of Physics, University of California, Berkeley, CA 94720, USA}
%\affiliation{Space Sciences Laboratory, University of California, Berkeley, CA 94720, USA}

\author{S. Kikuchi}
\affiliation{Yokohama National University, Yokohama, Kanagawa 240-8501, Japan}

\author{T. Kisner}
\affiliation{Computational Cosmology Center, Lawrence Berkeley National Laboratory, Berkeley, CA 94720, USA}

%\author{N. Krachmalnicoff}
%\affiliation{International School for Advanced Studies (SISSA), Via Bonomea 265, 34136, Trieste, Italy}

\author{A. Kusaka}
\affiliation{Physics Division, Lawrence Berkeley National Laboratory, Berkeley, CA 94720, USA}
\affiliation{Department of Physics, The University of Tokyo, Tokyo 113-0033, Japan}
\affiliation{Kavli Institute for the Physics and Mathematics of the Universe (Kavli IPMU, WPI), Berkeley Satellite, the University of California, Berkeley 94720, USA}
\affiliation{Research Center for the Early Universe, School of Science, The University of Tokyo, Tokyo 113-0033, Japan}

\author{A. T. Lee}
\affiliation{Department of Physics, University of California, Berkeley, CA 94720, USA}
\affiliation{Physics Division, Lawrence Berkeley National Laboratory, Berkeley, CA 94720, USA}

\author{D. Leon}
\affiliation{Department of Physics, University of California, San Diego, CA 92093-0424, USA}

\author[0000-0001-5536-9241]{E. Linder}
\affiliation{Space Sciences Laboratory, University of California, Berkeley, CA 94720, USA}
\affiliation{Physics Division, Lawrence Berkeley National Laboratory, Berkeley, CA 94720, USA}

\author{L. N. Lowry}
\affiliation{Department of Physics, University of California, San Diego, CA 92093-0424, USA}

%\author{A. Mangu}
%\affiliation{Department of Physics, University of California, Berkeley, CA 94720, USA}

\author[0000-0003-0041-6447]{F. Matsuda}
\affiliation{Kavli Institute for the Physics and Mathematics of the Universe (Kavli IPMU, WPI), UTIAS, The University of Tokyo, Kashiwa, Chiba 277-8583, Japan}
\author[0000-0001-9002-0686]{T. Matsumura}
\affiliation{Kavli Institute for the Physics and Mathematics of the Universe (Kavli IPMU, WPI), UTIAS, The University of Tokyo, Kashiwa, Chiba 277-8583, Japan}

\author[0000-0003-2176-8089]{Y. Minami}
\affiliation{Institute of Particle and Nuclear Studies (IPNS), High Energy Accelerator Research Organization (KEK), Tsukuba, Ibaraki 305-0801, Japan}

\author{M. Navaroli}
\affiliation{Department of Physics, University of California, San Diego, CA 92093-0424, USA}

\author[0000-0003-0738-3369]{H. Nishino}
\affiliation{Research Center for the Early Universe, School of Science, The University of Tokyo, Tokyo 113-0033, Japan}

\author[0000-0001-9396-8915]{A. T. P. Pham}
\affiliation{School of Physics, University of Melbourne, Parkville, VIC 3010, Australia}

\author[0000-0001-9807-3758]{D. Poletti}
\affiliation{International School for Advanced Studies (SISSA), Via Bonomea 265, 34136, Trieste, Italy}
\affiliation{Institute for Fundamental Physics of the Universe (IFPU), Via Beirut 2, 34151, Grignano (TS), Italy}
\affiliation{The National Institute for Nuclear Physics, INFN, Sezione di Trieste Via Valerio 2, I-34127, Trieste, Italy}

%\author[0000-0002-0689-4290]{G. Puglisi}
%\affiliation{Kavli Institute for Particle Astrophysics and Cosmology,SLAC National Accelerator Laboratory,2575 Sand Hill Rd, Menlo Park , CA 94025}

\author[0000-0003-2226-9169]{C. L. Reichardt}
\affiliation{School of Physics, University of Melbourne, Parkville, VIC 3010, Australia}

\author{Y. Segawa}
\affiliation{SOKENDAI (The Graduate University for Advanced Studies), Shonan Village, Hayama, Kanagawa 240-0193, Japan}
\affiliation{Institute of Particle and Nuclear Studies (IPNS), High Energy Accelerator Research Organization (KEK), Tsukuba, Ibaraki 305-0801, Japan}

%\author[0000-0001-7480-4341]{M. Silva-Feaver}
%\affiliation{Department of Physics, University of California, San Diego, CA 92093-0424, USA}

\author[0000-0001-6830-1537]{P. Siritanasak}
\affiliation{Department of Physics, University of California, San Diego, CA 92093-0424, USA}

%\author{N. Stebor}
%\affiliation{Department of Physics, University of California, San Diego, CA 92093-0424, USA}

%\author[0000-0002-9777-3813]{R. Stompor}
%\affiliation{AstroParticule et Cosmologie (APC), Univ Paris Diderot, CNRS/IN2P3, CEA/Irfu, Obs de Paris, Sorbonne Paris Cit\'e, France}

%\author[0000-0001-8101-468X]{A. Suzuki}
%\affiliation{Physics Division, Lawrence Berkeley National Laboratory, Berkeley, CA 94720, USA}

\author{O. Tajima}
\affiliation{Department of Physics, Kyoto University, Kyoto 606-8502, Japan}

\author[0000-0001-9461-7519]{S. Takakura}
\affiliation{Kavli Institute for the Physics and Mathematics of the Universe (Kavli IPMU, WPI), UTIAS, The University of Tokyo, Kashiwa, Chiba 277-8583, Japan}

\author{S. Takatori}
\affiliation{SOKENDAI (The Graduate University for Advanced Studies), Shonan Village, Hayama, Kanagawa 240-0193, Japan}
\affiliation{Institute of Particle and Nuclear Studies (IPNS), High Energy Accelerator Research Organization (KEK), Tsukuba, Ibaraki 305-0801, Japan}

\author{D. Tanabe}
\affiliation{SOKENDAI (The Graduate University for Advanced Studies), Shonan Village, Hayama, Kanagawa 240-0193, Japan}
\affiliation{Institute of Particle and Nuclear Studies (IPNS), High Energy Accelerator Research Organization (KEK), Tsukuba, Ibaraki 305-0801, Japan}

\author{G. P. Teply}
\affiliation{Department of Physics, University of California, San Diego, CA 92093-0424, USA}

\author{C. Tsai}
\affiliation{Department of Physics, University of California, San Diego, CA 92093-0424, USA}

\author[0000-0002-3942-1609]{C. Verg\`es}
\affiliation{Universit\'e de Paris, CNRS, Astroparticule et Cosmologie, F-75013 Paris, France}

\author[0000-0001-5109-9379]{B. Westbrook}
\affiliation{Department of Physics, University of California, Berkeley, CA 94720, USA}
\affiliation{Radio Astronomy Laboratory, University of California, Berkeley, CA 94720, USA}

\author[0000-0002-5878-4237]{Y. Zhou}
\affiliation{Department of Physics, University of California, Berkeley, CA 94720, USA}

\collaboration{1000}{(The \pb\ Collaboration)}

\begin{abstract}

We report a measurement of the \emode{} polarization power spectrum of the cosmic 
microwave background (CMB) using 150 GHz data taken from July~2014 to December~2016 with the \pb\ experiment. 
We reach an effective polarization map noise level of \mapdepth{} across an observation area of 670 square degrees. 
We measure the $EE$ power spectrum over the angular multipole range $500 \leq \ell  < 3000$, tracing the third to seventh acoustic peaks with high sensitivity. 
The statistical uncertainty on \emode{} bandpowers is $\sim$2.3\,$\mu {\rm K}^2$ at $\ell \sim 1000$ with a systematic uncertainty of 0.5\,$\mu {\rm K}^2$.  
The data are consistent with the standard \lcdm{} cosmological model with a probability-to-exceed of 0.38. 
We combine recent CMB  \emode{} measurements and make inferences about cosmological parameters in \lcdm{} as well as in extensions to \lcdm{}. 
Adding the ground-based CMB polarization measurements to the \planck{} dataset reduces the uncertainty on the Hubble constant by a factor of 1.2 to $H_0 = \lcdmCMBho{} ~{\rm km\,s^{-1} \,Mpc^{-1}}$. 
When allowing the number of relativistic species (\neff) to vary, we find $\neff{} = \neffCMBnnu$, which is in good agreement with the standard value of 3.046. 
Instead allowing the primordial helium abundance (\yhe) to vary, the data favor $\yhe{} = \yheCMByhe$. 
This is very close to the expectation of 0.2467 from Big Bang Nucleosynthesis. 
When varying both \yhe{} and \neff{}, we find $\neff = \yheneffCMBnnu$ and $\yhe{} = \yheneffCMByhe$.
\end{abstract}


\section{Introduction}%0.75p

Measurements of the cosmic microwave background (CMB) provide the foundation for our current understanding of cosmology. 
However, temperature measurements are now largely sample variance limited \citep{planck18-5} out to small angular scales where extragalactic foregrounds become significant \citep{george15,dunkley13,das14}. 
As a result, the focus of recent experiments has shifted to measuring the polarization of the CMB. 
CMB polarization anisotropies encode comparable amounts of information per angular multipole to the temperature anisotropy \citep{galli14}. Additionally, the relatively small polarization fraction of extragalactic sources \citep{gupta19, seiffert07,battye11} means that measurements can be extended to smaller angular scales before becoming foreground-dominated. 

The polarization patterns in the CMB are commonly separated into curl-free modes (\emodes) and gradient-free modes (\bmodes). 
This division is made because density fluctuations will produce \emodes{}, but not \bmodes{}, at first order. 
\bmodes{} are instead produced by gravitational waves and gravitational lensing \citep{seljak97, kamionkowski97b}. 

\emode{} anisotropy was first detected by DASI in 2002 \citep{Kovac_2002}. 
Since then, the field has moved from detecting power to high signal-to-noise ratio measurements of the power spectrum by a number of experiments \citep{planck18-5,louis17,bicepkeck18,henning18}. 
To date, these \emode{} measurements have supported the \lcdm{} cosmological model. 
Due to the lower levels of polarized foregrounds, polarization measurements have the potential to surpass the amount of information that can be extracted from the CMB temperature anisotropy, and thus improve our ability to constrain cosmological models \citep{galli14, louis17}.  
Measuring CMB polarization can also help disentangle effects that are degenerate in the temperature data.

In this paper, we report a measurement of the \emode{} auto-power spectrum ($EE$) in the angular multipole range, $500 \leq \ell  < 3000$, using new data collected between July 2014 and December 2016 from the \pb{} experiment. 
The expanded \pb{} survey covers 670 \sqdeg{} of sky at 150\,GHz, a 25-fold increase in area over the initial deep but small surveys by \pb{} \citep{polarbear17}. 
The survey region overlaps the \sptpol{} and \bicep{}2/\keckarray{} surveys, and the new \pb{} bandpowers provide an independent measurement of the \emode{} power spectrum on small angular scales.
A measurement of the \bmode{}  power spectrum on large angular scales on this field was presented by \citet[][hereafter PB19]{polarbear19}, which overlaps this work in the narrow range of angular scales $500 \leq \ell \leq 600$. 
%An upcoming paper will extend the \bmode{} power spectrum to $\ell=3000$. 
We combine the \pb{} bandpowers with other recent CMB power spectrum measurements \citep{planck18-5, louis17, story13} as well as CMB lensing power spectrum measurements \citep{planck18-8, wu19}, baryon acoustic oscillation (BAO) results \citep{beutler11, ross15,alam17} and Hubble constant measurements \citep{riess19} to study the implications for cosmology. 
This is the first time the cosmological implications of this combined dataset have been presented. 
%\citet[][hereafter PB19]{polarbear19}

This paper is organized as follows. 
In \S\ref{sec:instrument}, we give a brief overview of the \pb\ instrument and the 670\,\sqdeg{} survey. 
We continue to describe the low-level data processing and map-making in \S\ref{sec:todmaps}. 
The power spectrum analysis is outlined in \S\ref{sec:analysis}.
We test for systematic errors in \S\ref{sec:systematics}. 
In \S\ref{sec:bandpower}, we present the measurements of the E-mode power spectra. Subsequently, we study the cosmological implications in \S\ref{sec:cosmo}.
We conclude in \S\ref{sec:final}.

\section{The Polarbear 670 deg$^2$ survey }
\label{sec:instrument}

\pb{} is a receiver with 1274 cryogenically-cooled, transition-edge-sensor (TES) bolometers and a continuously rotating half-wave plate mounted on the 2.5\,m aperture Huan Tran Telescope at the James Ax Observatory on the Atacama plateau in Chile. 
The elevation (5190\,m) and the low Precipitable Water Vapor (PWV) of the Atacama plateau make the site one of the best in the world for microwave observations. 
Information of the instrument and telescope can be found in \citet{Arnold_SPIE2012, Kermish_SPIE2012,Takakura:2017ddx}. 

This work uses data taken with \pb{} on a 670\,\sqdeg{} field in three observing seasons from July 2014 to December 2016.
The field is centered at \patchloc, and largely overlaps the survey fields of  \bicep{}2/\keckarray{} \citep{bicepkeck18} and SPTpol \citep{henning18}. 
The data are taken in one-hour blocks by scanning back and forth at a constant velocity ($0 \fdg 4~\mathrm{s}^{-1}$) and constant elevation as the sky rotates past. 
After every four hours, the telescope is adjusted to track the field and the bolometers are retuned. 
More information on the scan strategy can be found in PB19. 

\section{Time-ordered data to maps}
\label{sec:todmaps}

In this section, we review the data selection and filtering of the time-ordered data (TOD). 
We then briefly describe the map-making process, and the determination of the beam function and absolute calibration. 
These steps closely follow the treatment in PB19, and we refer the reader to that work for more details while highlighting any differences from that work below.

\subsection{Data selection and filtering of the time-ordered data}\label{demod_dataselection}	

The data selection and filtering of the TOD are described in detail by PB19, and we repeat only the main points and differences opted due to the multipole range ($\ell \leq 600$ for PB19 and $500 \leq \ell  < 3000$ for this paper). %Briefly,
Periods of bad data, due to, e.g., weather or telescope turnarounds, are flagged and replaced by realizations of white noise before deconvolving the detector time constants and demodulating the effects of the continuously rotating HWP. 
The demodulated data is low-pass filtered and downsampled to 8 Hz (approximately $\ell<4000$) before being effectively high-pass filtered to reduce the impact of low-frequency noise by projecting out a ninth order polynomial from each subscan (which denotes one left going or right going motion of the telescope).\footnote{This was a first order polynomial in PB19, which sought to recover larger angular scales. The polynomial filter removes five times more modes in this work; however, the lower edge of the signal band has been increased by an even larger factor of ten (from $\ell=50$ to 500).}
The noise power spectral density for each TOD is fit to a model consisted of white noise and low frequency noise like in PB19. Detectors with unusually high or low noise levels at this point are flagged. 
Unlike PB19, we did not include cuts on the low-frequency noise performance here. 
%\tbd{Unlike PB19, the stacked timestream noise cuts are omitted because we don’t care about the low frequency common mode here. The map noise cuts are done on a $2^\prime$ pixel map created after a ninth-order polynomial filter and wafer-by-wafer common mode subtraction.}
A total of 3391 constant elevation scans (CESes) (each approximately one hour long) pass the cuts and are included in the analysis in this work. 

After data selection and demodulation, the TOD are filtered as follows. % filters are applied to the TOD after data selection. 
First, any significant, narrow-band instrumental lines, for instance due to electrical interference, are notch-filtered in Fourier space. 
%Fourth, to remove the slow temperature drift of the focal plane, a second-order polynomial is projected out from the entire CES timestream. 
Second, ground pickup is removed by subtracting a ground template separately from the $I$, $Q$, and $U$ TOD. 
Third, we estimate and subtract temperature-to-polarization leakage caused by detector non-linearity and telescope design through a principal component analysis (PCA), as demonstrated by \citet{Takakura:2017ddx}.
Note that the temperature-to-polarization leakage removal is only applied to real data and not the simulations in \S\ref{subsec:sims}. 
Fourth, as noted above, we project out a ninth order polynomial from each subscan. 
Finally, to reduce the effects of atmosphere, a common mode signal is straightly removed from all detectors while it is low-pass filtered before subtraction in PB19.

\subsection{Mapmaking} \label{mapmaking}

The cleaned TOD are binned into $2^\prime$ pixels, using the oblique Lambert equal area projection from a sphere to flat-sky. 
In this binning, the data are weighted according to each detector's power spectral density, which is consistent with white noise for individual detectors after filtering. 
 To simplify the power spectrum analysis, we combine the data from the set of 3391 CESes into 12 ``bundle'' maps that have relatively similar noise properties and map coverage. 
The effective map polarization noise level for fully combined data is \mapdepth{}, after we correct for the beam and transfer function of the filtering (see PB19). 

\subsection{Noise} \label{noisemodel}
Following PB19, we consider two noise models: sign-flip noise maps and simulated TOD noise realizations.
The sign-flip noise maps are created by randomly multiplying half of the CESes that enter a bundle map by $-1$, instead of $+1$, and thus nulling the true sky signal while maintaining the noise power. 
The simulated TOD noise consists of white noise plus low-frequency noise.
The TOD noise realizations are added to the simulated signal TOD to form simulated signal plus noise maps.
The sign-flip noise maps provide the fiducial estimate of the noise covariance for this work, with the TOD noise realizations being used to cross-check the results.
The TOD noise model is also used in the null test framework.

\subsection{Beams and calibration}
\label{subsec:beamcal}

The angular response of the instrument is determined using observations of Jupiter. 
As detailed by PB19, the beam is well-described by a Gaussian with a FWHM $3\farcm6$. 
The fractional uncertainty on the beam is determined by looking at the scatter in the recovered beam profile across the 50 Jupiter beam maps that pass quality cuts. 
Additionally, any errors in the pointing model will smear out the effective beam in the CMB survey maps. 
This pointing jitter is estimated by looking at bright sources in the survey region, and comparing the estimated FWHM on these sources to Jupiter. 
The beam uncertainties due to both the Jupiter measurements and jitter estimate are included in the likelihood as described in \S\ref{subsec:covariance}. 

The absolute gain calibration of the data is done in two steps. 
First, we determine the relative calibration between detectors so that their data can be coadded together into maps. 
The relative calibration of detectors is determined using a combination of a chopped thermal source (located at the secondary mirror) and Jupiter observations. 
Second, we compare the measured \emode{} power spectrum of these maps (see \S\ref{sec:bandpower}) to the predictions of the \planck{} best-fit \lcdm{} model to set the absolute calibration. 
While the latter step implicitly assumes isotropy across the sky, isotropy has already been stringently tested to better than the 2\% calibration uncertainty recovered in this work. 
One could get a much more precise calibration by comparing the actual temperature and polarization maps to \planck{} maps across this area (as was done by PB19), and thus eliminating the significant sample variance. 
However, we choose not to implement such a scheme since the calibration uncertainty does not limit the cosmological inferences of these data.

\section{Power spectrum analysis} \label{sec:analysis}%2 pages

The power spectrum is measured using a pseudo-$C_\ell$ cross-spectrum method \citep[][]{hivon02,tristram05}. 
The \pb{} implementation of this method has been previously described by \citet{polarbear14b}, and as ``Pipeline A'' by \citet{polarbear17} and PB19. 
In this section we outline the basic method while highlighting any changes from PB19. 
We express the bandpowers in terms of $D_\ell \equiv \ell (\ell + 1) C_\ell / (2\pi)$ unless otherwise noted.

Pseudo-$C_\ell$ methods are based on measuring the biased power spectrum, or pseudo-$C_\ell$,  from the fast Fourier transform (FFT) of an apodized map (or a spherical harmonic transform in curved sky), and then correcting these pseudo-$C_\ell$'s for the finite sky coverage, beams and filtering to recover the true spectrum on the sky. 
Cross-spectrum methods iterate on this approach by replacing auto-spectra by cross-spectra between maps with independent noise properties to avoid any noise bias. 

The binned pseudo-$C_\ell$'s can be written as:
%\begin{equation}
%\tilde{D}_b^{XY} = \frac{1}{\sum_{i^\prime \neq j^\prime} w_{X,i^\prime} w_{Y,j^\prime}}  \sum_{i \neq j} w_{X,i}w_{Y,j} \sum_{\mathbf{k} \in b} \frac{k(k+1)}{2\pi}\mathbf{\tilde{m}}_{\mathbf{k}}^{ X,i } \mathbf{\tilde{m}}_{\mathbf{k}}^{* Y,j} .
%\end{equation}
\begin{multline}
\tilde{D}_b^{EE} = \frac{1}{\sum_{i^\prime \neq j^\prime} w_{E,i^\prime} w_{E,j^\prime}} \times \\
 \sum_{i \neq j} w_{E,i}w_{E,j} 
\sum_{\mathbf{k} \in b} \frac{k(k+1)}{2\pi}\mathbf{\tilde{m}}_{\mathbf{k}}^{ E,i } \mathbf{\tilde{m}}_{\mathbf{k}}^{* E,j} .
\end{multline}
Here $w$ is a weight factor, and the indices $i$ and $j$ specify different bundle maps. 
The $\ell$-bin is denoted by $b$, the angular wave vector by $\mathbf{k}$, and the Fourier transform of an apodized bundle map by $\mathbf{\tilde{m}}_{\mathbf{k}}$. 
%Finally, $X/Y$ refer to $T$, $E$ or $B$. 
%As we only report \emode{} bandpowers in this work, we will simplify the equations in the remainder of the section by dropping explicit references to $X/Y$. 
%However, the pipeline tracks all the combinations. 

The true on-sky power spectrum is related to the  binned pseudo-$C_\ell$'s by:
\begin{equation}
D_b = \mathbf{K}_{bb'}^{-1} \tilde{D}_{b'}
\end{equation}
where the matrix $\mathbf{K}_{bb'}$ is known as the kernel matrix and defined by 
\begin{equation}
\mathbf{K}_{b b^{\prime}}=\sum_{\ell \ell^{\prime}} \mathbf{P}_{b \ell} \mathbf{M}_{\ell \ell^{\prime}} F_{\ell^{\prime}} B_{\ell^{\prime}}^{2} \mathbf{Q}_{\ell^{\prime} b^{\prime}}.
\end{equation}
Here, $\mathbf{P}$ and $\mathbf{Q}$ are binning and interpolation operators. 
The mode-coupling matrix $\mathbf{M}_{\ell \ell^{\prime}}$ accounts for the finite frequency resolution in the FFT of a finite area of sky. 
The beam function of the instrument is represented by $B_{\ell^{\prime}}$ (see \S\ref{subsec:beamcal}), while the transfer function $F_{\ell^{\prime}}$ accounts for the effects of filtering at the TOD and map levels. 
We will discuss these factors in more detail in the following subsections. %terms

\subsection{Apodization mask  and mode-coupling matrix}

We create an apodization mask in the following way. 
First we calculate the intersection of the non-zero weight regions of all 12 bundle maps. 
The edges of this region are smoothed by an $8^\circ$ Hamming window. 
We also mask bright radio sources, setting the mask to zero within a $10^\prime$ disk around each source, surrounded by a $10^\prime$ cosine taper. 
The maps are multiplied by this apodization mask and zero padded before being Fourier transformed. 

We calculate the mode-coupling matrix, $\mathbf{M}_{\ell \ell^{\prime}}$ for this apodization mask following the analytic expressions in Appendix A of \citet{hivon02}.

\subsection{Simulations} 
\label{subsec:sims}

We use end-to-end simulations to determine the filter transfer function in pseudo-$C_\ell$ methods as well as estimate the final bandpower uncertainties. 
We generate a suite of 192 simulated skies with an input signal drawn from the best-fit \lcdm{} model for TT,EE,TE+lowE+lensing in \citet{planck18-6}. 
The input skies are generated at a map pixel resolution of 1$^\prime$ and have only \emodes{}. 
%A second set of maps (not used in this work) has \bmodes-only. 
The simulated skies are re-observed using the real pointing information, and filtered exactly following the real data. One exception is the omission of the PCA filtering because temperature-to-polarization leakage is not added in simulated TODs.
We also run a subset (48) of these simulated skies through the null test framework to estimate the expected level of residual signal and scatter in each null test (see \S\ref{sec:systematics}).%(see \S\ref{subsec:null}). 
%As mentioned earlier, the measured temperature-polarization leakage is not subtracted from the simulated TODs. 

\subsection{Filter transfer function and bandpower window functions}

The transfer function, $F_{\ell}$, is calculated by comparing the \emode{} power spectrum of these simulations to the original input power spectrum as described in \citet{hivon02}.

We also report the bandpower window functions necessary to compare the binned spectra to a theory curve. 
In the pseudo-$C_\ell$ formalism, these bandpower window functions, $\mathbf{w}_{b\ell}$ can be expressed as:
\begin{equation}
\mathbf{w}_{b \ell}=\sum_{b^{\prime} \ell^{\prime}} \mathbf{K}_{b b^{\prime}}^{-1} \mathbf{P}_{b^{\prime} \ell^{\prime}}  \mathbf{M}_{\ell^{\prime} \ell} F_{\ell} B_{\ell}^{2}. 
\end{equation}
The bandpower window functions are applied to assumed theory spectrum, $ \mathbf{C}^{\rm theory}_\ell$, to get the binned expectation bandpowers,  
\begin{equation}
\mathbf{C}^{\rm theory}_b = \mathbf{w}_{b \ell} \mathbf{C}^{\rm theory}_\ell,
\end{equation}
for comparison with the measured bandpowers. 

We test the stability of the transfer function and bandpower window functions by running smaller numbers of simulations with different input cosmologies, and testing if the average resulting bandpowers for each simulated set match the expected bandpowers for the product of the bandpower window functions with the assumed cosmological model. 
We find agreement in all tests, validating the power spectrum pipeline.

\subsection{Bandpower Covariance}
\label{subsec:covariance}

We also need to estimate the uncertainty on the measured bandpowers. 
The total uncertainties will include sample and noise variance as well as the beam and calibration uncertainties. 
To allow the simulations to be run before settling on the final absolute calibration, we calculate the sample and noise variance separately before combining the two estimates. 

We use the 192 mock-observed noiseless CMB maps from \S\ref{subsec:sims} to estimate the covariance matrix due to sample variance. 
We use the calibrated, sign-flip noise maps from \S\ref{mapmaking} to estimate the noise variance, while cross-checking the results with the simulated noise maps. 
For both the sample and noise variance, we estimate the covariance matrix at an initial binning of $\Delta \ell  =50$, and condition this matrix following \citet{henning18} to reduce the impact of uncertainties in the covariance estimate. 
Specifically, we require the correlation matrix  to be a symmetric Toeplitz matrix. 
Given the expected correlation length, we also zero out the correlation for $\Delta\ell > 150$. 
The observed correlation at these $\Delta\ell$s is consistent with zero (although the uncertainty is large). 
We then rebin this estimate of the sample variance into the final bandpower binning.

Beam and calibration uncertainties are dealt with separately. 
We handle the calibration uncertainty by adding a calibration factor to the cosmological analysis with a prior set by the expected 2\% calibration uncertainty. 
The beam uncertainty is propagated into a beam correlation matrix, $\rho_{b b'}$. 
At each step in the chain, this beam correlation matrix is combined with the binned theory spectrum $D_b$ and added to the sample and noise covariance matrix to yield the total covariance at that step: 
\begin{equation}
C^{\rm tot}_{bb'} = C_{bb'}^{\rm S+N} + \rho_{b b'} D_b D_{b'} 
\end{equation}

\section{Data validation} \label{sec:systematics}

We test the data for unknown systematics using null tests. 
Each null test splits the dataset in approximately half, with the splits chosen to be sensitive to likely sources of systematic bias. 
The difference between the two halves removes nearly all true sky signal, thus suppressing the sample variance and allowing a more sensitive test for systematics.
As will be described in more detail below, the null test suite shows no evidence for systematics in the data.

We have also run a suite of simulations for expected sources of systematic errors (information on the simulation procedure can be found in PB19). 
At $\ell>1050$, the most significant systematics are related to detector cross-talk, pointing, and the half-wave plate; the estimated systematic uncertainty is less than 0.25 the statistical uncertainty in all bins  (\pb{} Collaboration, in prep.). 
Given that the systematic uncertainties are small compared to the statistical uncertainties on the \emode{} bandpowers, we choose to neglect the systematic errors in this work. 
%We conclude that systematic errors are not significant for the \emode{} bandpowers. 

%\subsection{Null Tests}\label{subsec:null}

We run a suite of 19 null tests to search for potential bias in our data set. 
Our framework has been previously used by \citet{polarbear14b,polarbear17}, and PB19, which is based on the formalism developed originally by the \quiet\ Collaboration \citep{PhDT_Bischoff}.
The binned null spectrum, $\hat{C}_{b}^{\rm null}$, is constructed as:
\begin{equation}
\hat{C}_{b}^{\rm null} = \hat{C}_b^{A} + \hat{C}_b^{B} - 2\hat{C}_b^{AB},
\end{equation}
where $\hat{C}_b^{A/B}$ are the spectra calculated following \S\ref{sec:analysis} for each half of the data split, and  $\hat{C}_b^{AB}$ is the cross-spectrum between the two halves (all after correcting for the appropriate filter transfer functions and mode-coupling matrices). 
For each null test, the data from each half is re-bundled to maximize the overlapping area. 
The binning in $\ell$ used in these null tests is the same as in Table \ref{spectrumtable}. 

Most of these null tests have been previously described by PB19, but five are added to test specific potential concerns for the \emode{} measurement. 
The new tests include: (1) a second test on Sun contamination, splitting the data by the distance to the Sun; (2) a test splitting the data based on the observed level of temperature-to-polarization leakage in each CES; (3 \& 4) two tests of HWP contamination by splitting the data on the level of either the $2f$ or $4f$ line amplitude in each CES; and (5) a random split of the bolometers to test the quality of the noise model. 
A complete description of 19 tests can be found in Appendix.
We estimate the uncertainty on each null test bin, $\sigma (\hat{C}_b^\mathrm{null})$, by looking at the standard deviation of a suite of 48 simulated null spectra. 
We then define the statistic, 
\begin{equation}(\chi_b^\mathrm{null})^2 \equiv \left(\frac{\hat{C}_b^\mathrm{null} }{\sigma (\hat{C}_b^\mathrm{null}) }\right)^2.
\end{equation}
We compare the values of  $(\chi_b^\mathrm{null})^2$ from the real data to simulations to calculate the PTE for each test.  
Summing across all tests and all bins, we find the PTE for the total $\chi^2$ to be 67.9\%. 
The data thus show no evidence for systematic biases. 

We also test that the set of PTEs is consistent with a uniform distribution as expected. 
Specifically, we perform a Kolmogorov-Smirnov (KS) test on the three sets of PTEs of the $\chi^2_\mathrm{null}$ values by test, by bin, and overall. 
All three distributions are consistent with uniform distributions (PTE = 0.45, 0.30, 0.13), showing no evidence for a bias. 

\comment{
\subsection{Simulation of Systematics} \label{systematicspipeline} \tbd{update}

In parallel with the null tests, we evaluate systematic errors from several known sources including ones mentioned in Section~\ref{calibration}. The pipeline and the systematic effects have been described in PB BB19. In brief, we create 4 simulated set of maps through our simulation map making pipeline and insert the bias effect at the TOD level for each systematic. We use our cross spectrum pipeline on the maps, then define the systematic error as the mean of the differences between power spectra of maps that contain systematics and maps that are bias-free for each case.  The error from pointing is noticed to be the main contributor up to $\ell = 1800$ due to sample variance. However, we find the total bias added from all effects to be secondary to the statistical error. We hence only list the numerical values in Table \ref{spectrumtable} along with our bandpowers.}

\section{Bandpowers} \label{sec:bandpower}%- 0.5p text + couple of big plots

\begin{figure}
\begin{center}%[b]{0.5\textwidth}

\includegraphics[width=0.49\textwidth]{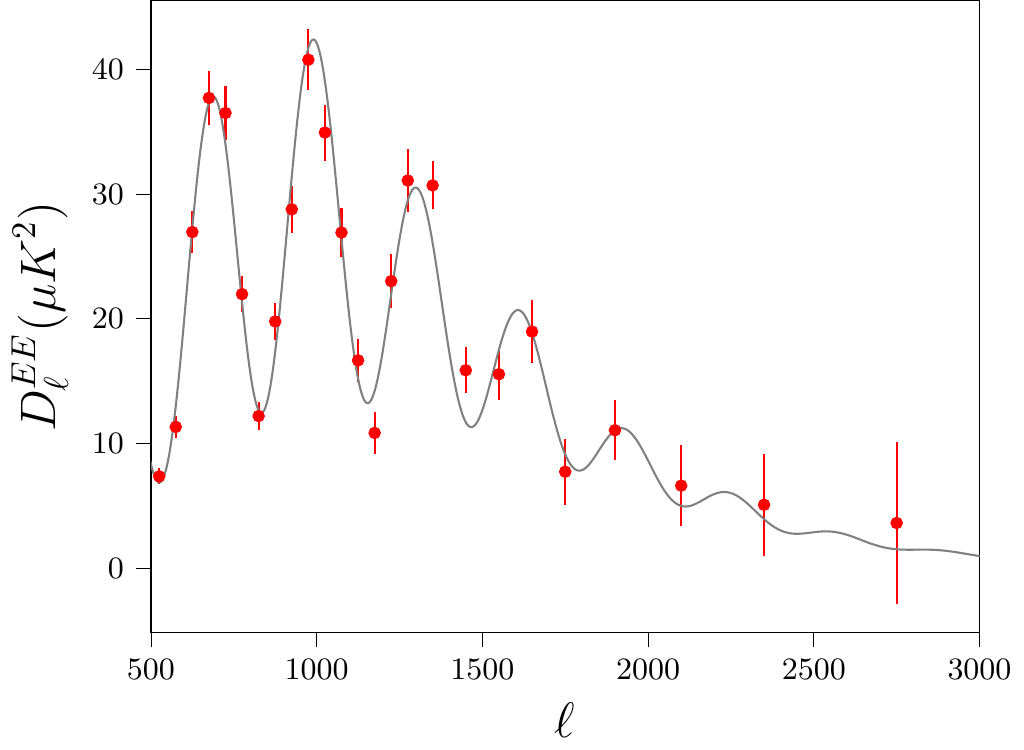}
%\resizebox{.5\linewidth}{!}{\input{real_figures/ee_show.tex}}
%\input{real_figures/EE_show.tex}
\end{center}
\caption{Measured \pb{} $E$-mode spectrum with included error bars from statistical uncertainties. The solid gray line represents the best-fit \lcdm{} model from \citep{planck2015XIII}. The bandpowers and statistical errors are listed in Table~\ref{spectrumtable}.}
\label{fig_eebp}
\end{figure}

The \emode{} bandpowers measured by applying the analysis method of \S\ref{sec:analysis} to the \pb{} 670\,\sqdeg{} survey are shown in Figure \ref{fig_eebp} and tabulated in Table \ref{spectrumtable}. 
\emode{} power is detected at very high significance, with zero \emode{} power excluded at 61\,$\sigma$. 
The \pb{} bandpowers are consistent with the \lcdm{} model; the \pb{} data has a PTE of 0.38 relative to the best-fit \lcdm{} model for the \pb{} and \planck{}  \citep{planck18-5} datasets. 
The \pb{} bandpowers trace out the third through seventh acoustic peaks in the \emode{} spectrum, and extend to $\ell=3000$ well into the Silk damping tail \citep{Silk68}.  

\begin{figure*}
\begin{center}
\includegraphics[width=1.\textwidth]{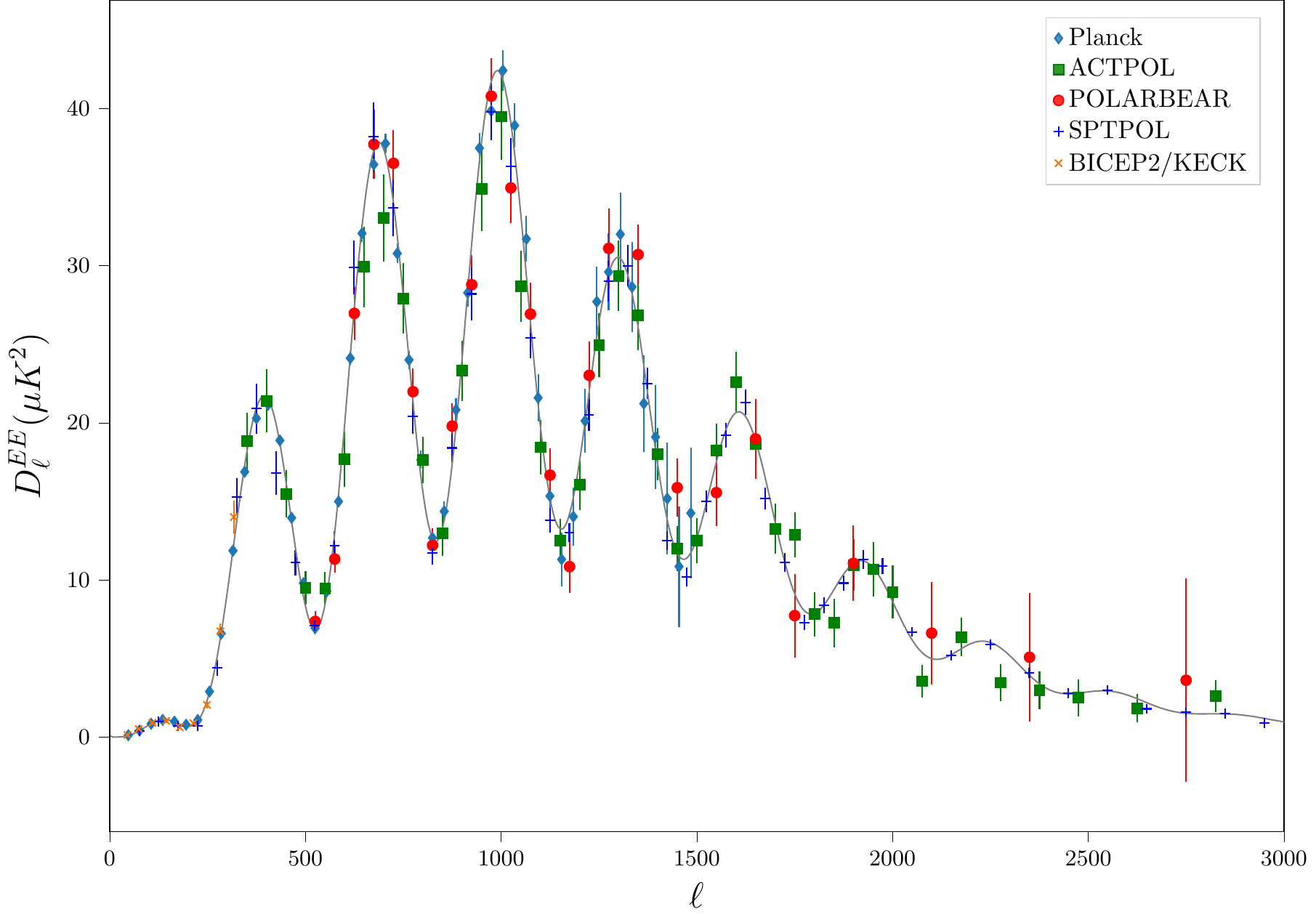}

\end{center}
\caption{Recent CMB \emode{} power spectrum measurements from \planck{} \citep{planck18-5}, \pb{}, \sptpol{} \citep{henning18}, \actpol{} \citep{louis17}, and BICEP2/Keck \citep{bicepkeck18} have mapped out the \emode{} power spectrum at high S/N from very large scales out to the ninth acoustic peak. 
The power spectrum measurements from the different experiments are in good agreement with each other. 
The data used in the parameter section of this work are shown by filled points. The solid gray line represents the best-fit \lcdm{} model from \citep{planck2015XIII}. }
\label{fig:currentEE}
\end{figure*}

We show the current state of \emode{} power spectrum measurements in Figure \ref{fig:currentEE}. 
In this figure, we compile the bandpowers of this work with other recent \emode{} measurements \citep{louis17, planck18-5, bicepkeck18, henning18}. 
The observed \emode{} spectra agree well, enhancing our confidence in the \emode{} measurements.

% & ~$D_\ell^{EE}$ cal error $(\mu K^2)$~ 
\begin{table}
\begin{center}

\caption{Bandpowers}  \label{spectrumtable}
%\resizebox{\columnwidth}{!}{%
\begin{tabular}{ c c c c   }
\hline
Multipole range & ~ $\ell_{\rm eff}$~ & ~$D_b^{EE}$ $[\mathrm{\mu K}^2]$~ & ~$\sigma(D_b^{EE})$  $[\mathrm{\mu K}^2]$~ \\
\hline
$500 \leq \ell <550$ &	525.4 & 7.36	 & 0.64	 \\
$550 \leq \ell <600$ &	575.3 & 11.33& 0.87	\\
$600 \leq \ell <650$ &	625.3 & 26.96	& 1.68 		  \\
$650 \leq \ell <700$ &	675.2 & 37.72	& 2.19		  \\
$700 \leq \ell <750$ &	725.4 & 36.51	& 	2.13	\\
$750 \leq \ell <800$ &	775.2 & 21.98	& 1.46	 \\
$800 \leq \ell <850$ &	825.3 & 12.20	& 1.09  \\
$850 \leq \ell <900$ &	875.2 & 19.79	& 1.45 	 \\
$900 \leq \ell <950$ &	925.3 & 28.79	& 1.87	\\
$950 \leq \ell <1000$ &	 975.2 & 40.78	& 2.43	 \\
$1000 \leq \ell <1050$ &	1025.3 & 34.95	 & 2.25  \\
$1050 \leq \ell <1100$ &	1075.2	& 26.92	 & 2.00	\\
$1100 \leq \ell <1150$ &	1125.2	& 16.67 & 1.72  \\
$1150 \leq \ell <1200$ &	1175.1	& 10.89 & 1.65  \\
$1200 \leq \ell <1250$ &	1225.2	& 23.02 & 2.16	\\
$1250 \leq \ell <1300$ &	1275.1	& 31.09	 & 2.55 \\
$1300 \leq \ell <1400$ &	1350.6	& 30.71 & 1.93	 \\
$1400 \leq \ell <1500$ &	1450.6	& 15.88	 & 1.84 \\
$1500 \leq \ell <1600$ &	1550.6	& 	15.56 & 2.11  \\
$1600 \leq \ell <1700$ &	1650.5	& 18.98 & 2.54	 \\
$1700 \leq \ell <1800$ &	1750.5	& 7.74& 2.65	\\
$1800 \leq \ell <2000$ &	1901.8	& 11.06 & 2.40  \\
$2000 \leq \ell <2200$ &	2101.6	& 6.62 & 3.27 \\
$2200 \leq \ell <2500$ &	2353.2	& 5.08 & 4.08 \\
$2500 \leq \ell <3000$ &	2757.6	& 3.62	& 6.47\\
\hline
\end{tabular}  \par
%}
\end{center}
\tablecomments{The angular multipole range,  \emode{} bandpowers, and uncertainties for the \pb{} survey. 
The uncertainties are taken from the square root of the diagonal of the covariance matrix, and do not include beam or calibration errors. }
\end{table}

\section{Cosmological implications} \label{sec:cosmo}%3p

We now turn to the cosmological implications of the \pb{} \emode{} power spectrum along with other recent cosmological observations. 
We look at parameter constraints for the standard, six-parameter, \lcdm{} cosmological model. 
We also look at two one-parameter extensions to \lcdm{}, \neff{} or \yhe.
These extensions are constrained primarily by the Silk damping scale in temperature data. 
Finally, we consider the two-parameter extension of \yhe+\neff, of interest as a test of big bang nucleosynthesis (BBN).

\subsection{Methodology}

We derive parameter constraints using the 2019 version of the Markov Chain Monte Carlo (MCMC) package \textsc{CosmoMC} \citep{lewis02b}. 
We have extended \textsc{CosmoMC}  to include the \pb{} bandpowers in a manner similar to the public likelihood for \citet{henning18}. 
The \pb{} likelihood code and associated data are available on the LAMBDA website.

In addition to the usual cosmological parameters in \textsc{CosmoMC}, we have added four nuisance parameters specific to the \pb{} data, most with an informative prior. 
The first parameter is the calibration factor (in power) for the \pb{} \emode{} power spectrum. 
We set a prior on this factor based on the expected 2\% uncertainty in the absolute calibration. 
The other three parameters relate to on-sky signals. 
First, we have one term to describe the polarized Poisson-distributed point source power in the field after masking. 
This power scales with $\ell$ as $D_\ell \propto \ell^2$, and we report the power at $\ell=3000$, $D_{3000}^{PS}$.
We use a weakly informative prior on the point source power that is uniform for $D_{3000}^{PS} \in [0,10]\,$\uksq. 
Second, we have one parameter to describe polarized Galactic dust, which we model as having a power spectrum,
\begin{equation}
D_\ell^{\rm dust} = D_{80}^{\rm dust} \left(\frac{\ell}{80}\right)^{\alpha_{\rm dust}}.
\end{equation}
%_{\rm dust}
Given that the \pb{} bandpowers are at much higher angular multipoles, we apply strong priors from \citet{bicepkeck18}. 
Specifically, we fix $\alpha_{\rm dust}=-0.58$ and apply a Gaussian prior that $D_{80}^{\rm dust}$ is drawn from $N(0.0188, 0.0042^2)\,\uksq$. 
The results are insensitive to this term; we have run one chain for \lcdm{}+\neff{} with the dust power zeroed and have seen no shifts larger than $0.1\,\sigma$. 
Finally, we allow for ``super-sample lensing'' variance \citep{manzotti14}. 
This is parametrized by the mean lensing convergence across the field, $\kappa$, to which we apply a Gaussian prior centered at zero with a 1\,$\sigma$ width of 0.001.\footnote{We estimate the prior width of 0.001 for the \pb{} survey area from Fig.~2 of \citet{manzotti14}.}

\subsection{Data sets}

We include the \planck{} 2018 TT, TE, and EE power spectra likelihoods in all results \citep{planck18-5}. 
Constraints from \planck{} alone are referred to by `\planck'. 
We also explore the effects of adding ground-based CMB measurements, specifically the SPT-SZ TT measurements \citep{story13}, ACTpol TT/TE/EE measurements \citep{louis17}, and the \pb{} EE spectrum in this work. 
We do not include  \bicep{}2/\keckarray{} data at large angular scales as we do not look at the tensor-to-scalar ratio. 
We include the ACTpol (but not SPTpol \citep{henning18})  TE/EE measurements because the ACTpol survey region does not overlap the \pb{} survey while the \sptpol{} survey has nearly 100\% overlap. 
 Accounting for the common sample variance would be a non-trivial exercise and is not possible using only the publicly available likelihood. 
Given the  agreement between the \emode{} measurements in Fig.~\ref{fig:currentEE}, we are confident that it is appropriate to combine these different datasets. 
Constraints from the combination of \planck{} 2018 and the ground-based CMB measurements are referred to by `CMBselect'. 

We also consider what impact data besides the primary CMB power spectra have on the cosmological constraints. 
Here we include the lensing power spectra from \planck{} and SPTpol \citep{planck18-8, wu19}. 
We also include the \citet{riess19} local measurement of the Hubble constant, $H_0 = 74.03 \pm 1.42$\,km/s/Mpc. 
Lastly, we include three baryon acoustic oscillation measurements: the SDSS-III BOSS DR12 Consensus sample \citep{alam17}, the DR7 MGS sample \citep{ross15} and the 6dFGS survey \citep{beutler11}. 
We label constraints that include these data in addition to the CMB power spectrum data by `CMBext'.

\subsection{Constraints on the \lcdm{} model} \label{fitCDM}

As has been previously noted, the \planckeight{} CMB power spectrum data alone do an excellent job of constraining all six parameters in the standard \lcdm{} model. 
We report the median parameter values and 68\% confidence intervals in Table~\ref{table_lcdm}. 
While the optical depth is relatively uncertain with a 14\% error bar, the other five parameters are measured with percent-level precision. 
Adding the ground-based CMB power spectrum measurements to the set reduces the allowed parameter volume by a factor of 2.0, or roughly a 7\% reduction in parameter uncertainties.  %the cmbext one is a factor of 5.1, although not quoted
These small-scale power spectrum measurements do not help recover the optical depth however, as that parameter constraint depends on the reionization bump at $\ell < 10$. 
Adding the CMB lensing, BAO, and \ho{} data yields a further reduction in uncertainties of order 10\%. 
Notably, lensing provides another avenue to measure the amplitude of fluctuations, $A_s$, which partially breaks the $A_s e^{-2\tau}$ degeneracy in the power spectrum alone and helps the determination of $A_s$ and $\tau$ individually. 

There has been much discussion recently about the degree of tension between local determinations of the Hubble constant, \ho{}, and the values inferred from the \planck{} data \citep[e.g.,][]{riess19,wong19}. 
Adding the ground-based CMB \emode{} bandpowers to the \planck{} data supports the current tension on the Hubble constant by reducing the uncertainty by a factor of 1.2 to:
\begin{equation}
H_0 = \lcdmCMBho{} ~{\rm km\,s^{-1} \,Mpc^{-1}}. 
\end{equation}
As a side note, we see a similar level of improvement when adding only the \pb{} bandpowers to \planck{} (a factor of 1.13 reduction in uncertainty on \ho{}) or adding only the other ground-based CMB bandpowers (a factor of 1.19). 
The tension with the local determination by \citet{riess19} of $H_0 = 74.03 \pm 1.42~{\rm km\,s^{-1} \,Mpc^{-1}}$ remains essentially unchanged from 4.3 to $4.5\,\sigma$.

%we didn't turn on sigma8, so can't discuss changes...

\begin{table*}
\begin{center}
\caption{\lcdm{} parameter constraints }  
\begin{tabular} { c | c  c c}

  &  Planck & CMBall & CMBext\\
\hline\hline
{\boldmath$\Omega_b h^2   $} & \lcdmPlanckombh{}  & \lcdmCMBombh{} & \lcdmCMBextombh{}\\

{\boldmath$\Omega_c h^2   $} & \lcdmPlanckomch{}  & \lcdmCMBomch{} & \lcdmCMBextomch{} \\

{\boldmath$100\theta_{MC} $} & \lcdmPlancktheta{} &  \lcdmCMBtheta{} & \lcdmCMBexttheta{} \\

{\boldmath$\tau           $} & \lcdmPlancktau{}  &  \lcdmCMBtau{} & \lcdmCMBexttau{} \\

{\boldmath${\rm{ln}}(10^{10} A_s)$} & \lcdmPlancklogA{} &  \lcdmCMBlogA{} & \lcdmCMBextlogA{} \\

{\boldmath$n_s            $} & \lcdmPlanckns{} & \lcdmCMBns{} & \lcdmCMBextns{} \\
\hline
 {\boldmath $H_0$} (km s$^{-1}$ Mpc$^{-1}$) & \lcdmPlanckho{} & \lcdmCMBho{} & \lcdmCMBextho{} \\
\hline\hline
\end{tabular} 
\label{table_lcdm}
\tablecomments{The median values and 68\% confidence intervals for the six \lcdm{} parameters for the \planck, CMBall and CMBext datasets. 
Adding the ground-based CMB power spectrum measurements to \planck{} reduces the parameter uncertainties by $\sim$7\% (except for the amplitude terms $A_s$ and $\tau$ which depend on the low$-\ell$ polarization bump). 
Adding the  BAO,  \ho{} and CMB lensing data further reduces uncertainties by an additional $\sim$10\%. }
\end{center}
\end{table*}

\begin{figure}
\begin{center}
\includegraphics[width=0.48\textwidth]{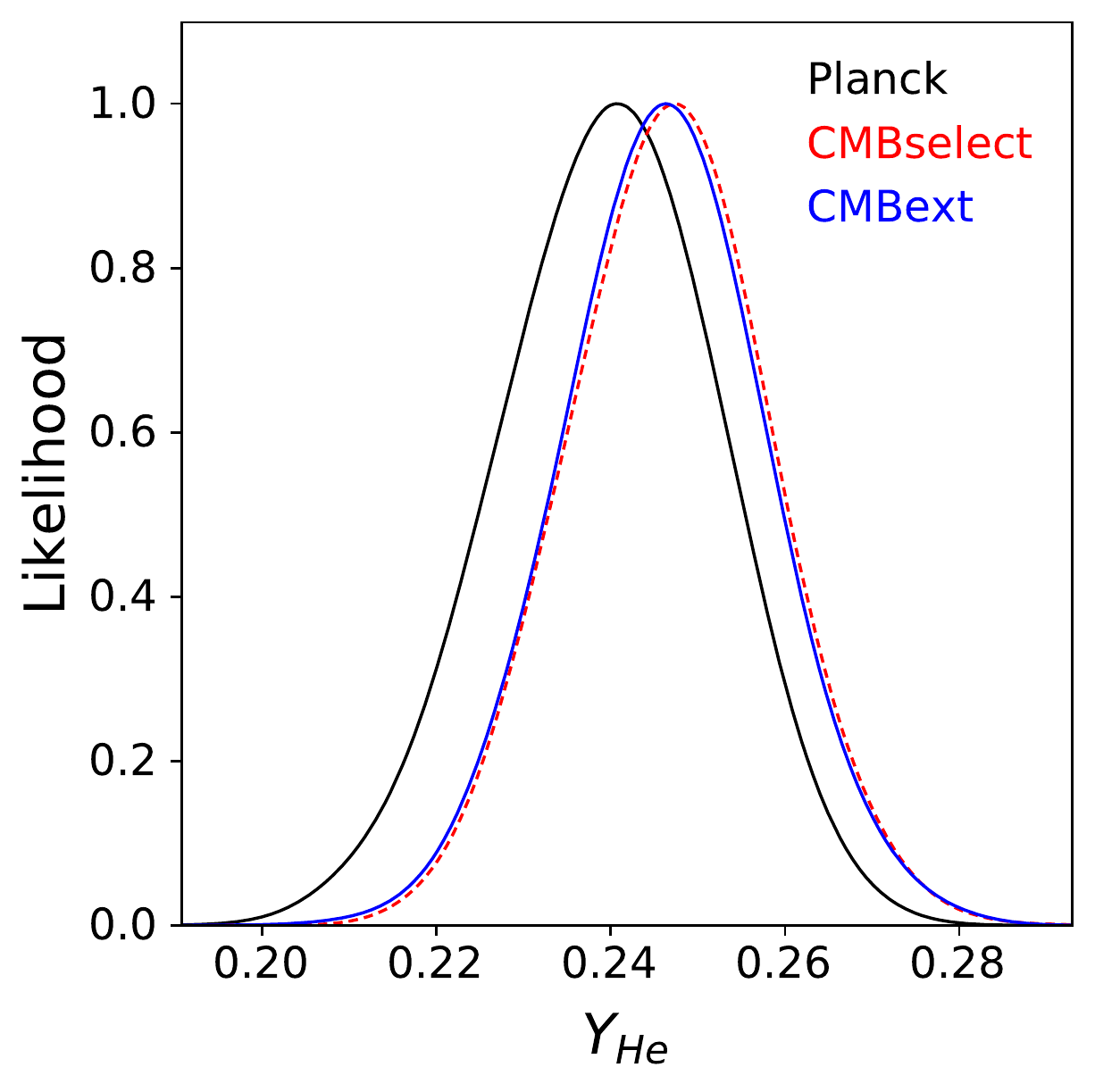}
\end{center}
\caption{\label{fig_yhepro}  Posterior probability distribution function for the primordial helium abundance, \yhe{}, for the three datasets. 
Adding the ground-based CMB data to \planck{} slightly shifts the preferred helium abundance towards the BBN prediction of $\yhe \sim 0.2467$, but leads to only a minor reduction in the uncertainty. 
All three datasets are in good agreement with the BBN prediction. }
\end{figure}

\subsection{Constraints on the primordial helium abundance}

We also look at the inferred primordial helium fraction, which can be viewed as a test for new particles or physics during the epoch of big bang nucleosynthesis (BBN). 
The expected helium fraction under BBN consistency in the \lcdm{} model for the CMBselect dataset is extremely tightly constrained at $\yhe  = \lcdmCMBypbbn{}$. 
Relaxing BBN consistency substantially weakens what we can infer about the helium fraction. 
However, the CMB anisotropies have some sensitivity to the helium fraction as it changes the number of free electrons preset at recombination. 
Higher helium fractions lead to fewer free electrons, a longer photon mean free path and thus more Silk damping. 
Using \planck{} alone, we find $\yhe  = \yhePlanckyhe{}$. 
Adding the other CMB measurements improves this slightly to 
\begin{eqnarray}
\yhe  &=& \yheCMByhe{},
\end{eqnarray}
in excellent agreement with the expectation from BBN. 
There is no further improvement from adding the other cosmological data. 

%H0_Planck=$67.19\pm 0.69$
%H0_allCMB = $67.26\pm 0.67$
%H0_CMBBAO=$67.42\pm 0.60$

\subsection{Constraints on the number of relativistic species}

The energy density of relativistic particles in the early Universe is proportional to \neff, the effective number of relativistic species. 
The standard model of particle physics predicts that $\neff = 3.046$ for the three neutrino species plus a small correction from positron annihilation \citep{Mangano_2005}. 
The preferred \neff{} from the CMBselect dataset is within $0.4\,\sigma$ of this prediction:
\begin{equation}
\nonumber \neff = \neffCMBnnu.
\end{equation}
Adding the non-CMB data (i.e. the CMBext dataset) slightly reduces the preferred value of \neff, but it remains within $0.9\,\sigma$ of the expectation:
\begin{equation}
\nonumber \neff = \neffCMBextnnu.
\end{equation}

\begin{figure*}
\begin{center}
\includegraphics[width=0.9\textwidth]{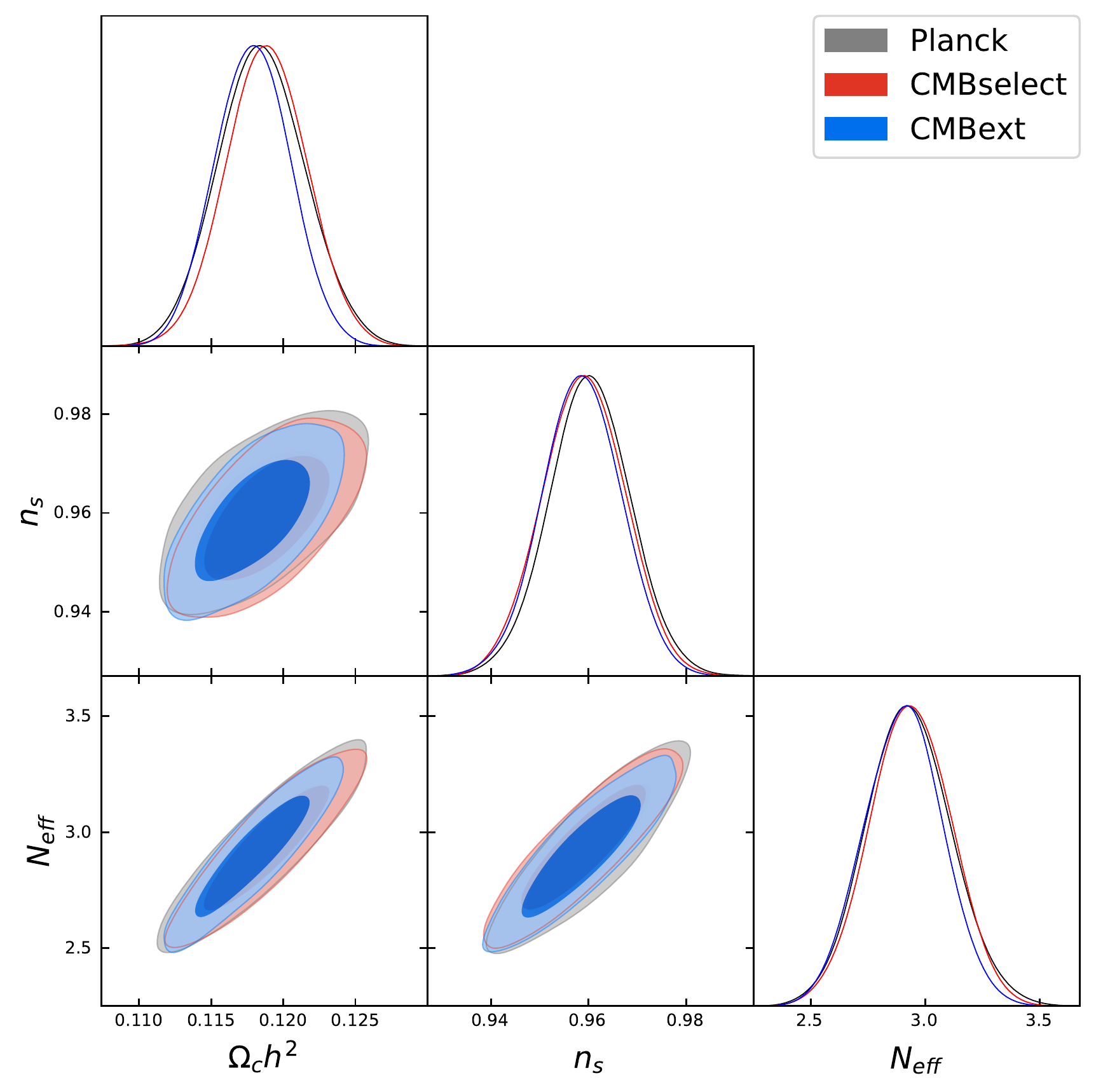}
\end{center}
\caption{Posteriors for the parameter subset  \neff, $n_s$, and $\Omega_c h^2$ for the \planck, CMBselect, and CMBext datasets. 
Adding data beyond the \planck{} CMB bandpowers only modestly reduces the allowed parameter volume without significantly shifting the preferred values or breaking the parameter degeneracies. }
\label{fig_neff}
\end{figure*}

When \neff{} is allowed to vary, as shown in Figure~\ref{fig_neff} we see that it correlates strongly with $\Omega_c h^2$ and $n_s$. 
As discussed by \citet{hou13}, increasing the matter density as \neff{} increases avoids shifting the redshift of matter-radiation equality. 
A side effect is that the CMB constraint on the Hubble constant significantly weakens: the uncertainty on the Hubble constant nearly triples from $\pm 0.57$ to $\pm 1.5$ km s$^{-1}$ Mpc$^{-1}$ (the central value changes by less than $0.4\,\sigma$ of the weakened constraint).

%H0_Planck=$66.5\pm 1.4$
%H0_allCMB=$66.4\pm 1.4$
%H0_CMBBAO=$66.4\pm 1.3$

\comment{
\begin{table*}
\begin{center}
\caption{Parameter constraints for \lcdm\ +$N_{Eff}$}  \label{table_neff}
\hskip-4.0cm\begin{tabular} { l  c  c c}

 Parameter &  Planck   & CMB & CMBext \\
\hline
{\boldmath$\Omega_b h^2   $} & $0.02226\pm 0.00022 $ & $0.02218\pm 0.00021 $ & $0.02220\pm 0.00021 $\\

{\boldmath$\Omega_c h^2   $} & $0.1185\pm 0.0030 $ & $0.1188\pm 0.0028 $& $0.1179\pm 0.0026$\\

{\boldmath$100\theta_{MC} $} & $1.04109\pm 0.00045 $  & $1.04117\pm 0.00040 $& $1.04124\pm 0.00039$\\

{\boldmath$\tau           $} & $0.0538^{+0.0071}_{-0.0082}$  & $0.0508\pm 0.0076 $ & $0.0499\pm 0.0078$\\

{\boldmath$N_{eff}        $} & $2.93\pm 0.18 $  & $2.94\pm 0.18 $ & $2.90\pm 0.17$\\

{\boldmath${\rm{ln}}(10^{10} A_s)$} & $3.040^{+0.017}_{-0.019} $  & $3.036\pm 0.018  $ & $3.031\pm 0.017 $\\

{\boldmath$n_s            $} & $0.9603\pm 0.0083 $ & $0.9590\pm 0.0084 $ & $0.9586\pm 0.0081          $\\
\hline
\end{tabular} \par
\end{center}
\vspace{8pt}
\noindent Values of 6 parameters 68\% limits from the constraint of  \lcdm\ +$N_{eff}$ with 3 combinations of datasets.
\end{table*}}

\subsection{Constraints on the \lcdm{}+\yhe{}+\neff{} model}

\begin{figure*}
\begin{center}
\includegraphics[width=1.\textwidth]{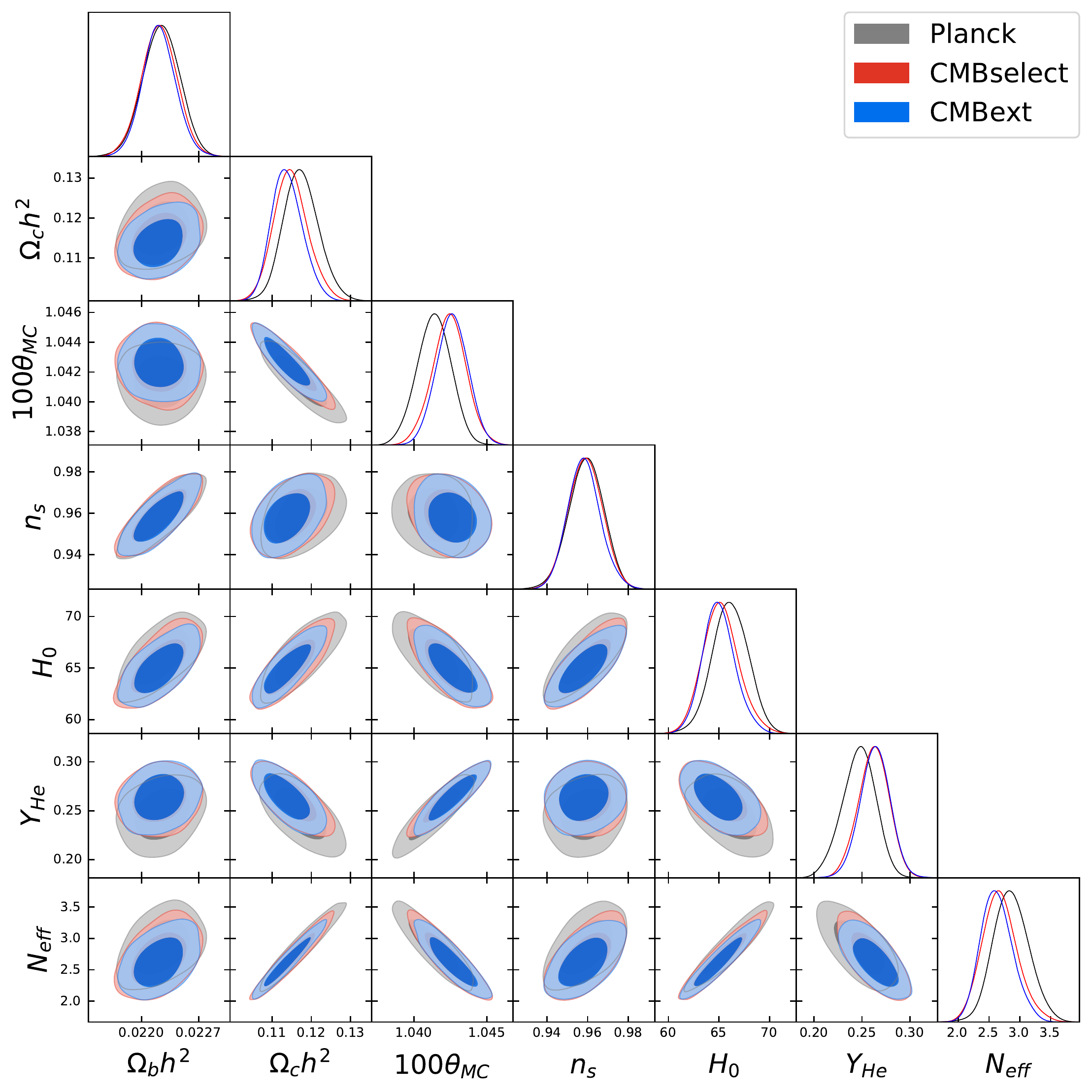}
\end{center}
\caption{Parameter posteriors for the \lcdm+\yhe+\neff{} model. 
We have excluded the optical depth $\tau$ and amplitude of scalar perturbations $A_s$ to reduce the complexity of the figure as these two parameters change negligibly between the datasets. }
\label{fig_yheneff}
\end{figure*}

We now consider the results when allowing both \neff{} and \yhe{} to vary, as both parameters affect the damping tail. 
As with the other extensions to \lcdm{} considered, freeing these two parameters does not significantly improve the quality of the fit ($\Delta\chi^2 \simeq -1.2$ for two new parameters for the CMBext dataset). 
The resulting parameter posteriors are shown in Figure~\ref{fig_yheneff}.
We find for the CMBselect dataset:
\begin{eqnarray}
\nonumber \neff &=& \yheneffCMBnnu,\\
\nonumber \yhe &=& \yheneffCMByhe. 
\end{eqnarray}
The CMBext dataset prefers essentially the same values as well: 
\begin{eqnarray}
\nonumber \neff &=& \yheneffCMBextnnu,\\
\nonumber \yhe &=& \yheneffCMBextyhe.
\end{eqnarray}

Adding the ground-based CMB measurements of the damping tail to the \planck{} bandpowers pushes along the \neff{}/\yhe{} degeneracy towards higher values of \yhe, and lower values of \neff. 
However, Fig.~\ref{fig_yheneff} shows that the $2\,\sigma$ parameter ellipses still contain the \lcdm{} values and, as mentioned above, the quality of the fit does not substantially improve. Table \ref{table_cmbext} summarizes the median and 68\% confidence intervals for the parameters in the \lcdm{} case and extensions, for the CMBext dataset. 

%H0_Planck=$66.1\pm 1.8$
%H0_allCMB= $65.2^{+1.6}_{-1.9}$
%H0_CMBBAO=$65.0^{+1.4}_{-1.7}$

\begin{table*}
\begin{center}
\caption{Parameter constraints for the CMBext dataset} 
%\hskip-4.0cm
\begin{tabular} { c | c c c c c}

 & \lcdm{} & \lcdm{}+\neff{} & \lcdm{} +\yhe{} & \lcdm{}+\yhe{}+\neff{}\\
\hline\hline
{\boldmath$\Omega_b h^2 $} & \lcdmCMBextombh{} &  \neffCMBextombh{} & \yheCMBextombh{} & \yheneffCMBextombh{} \\

{\boldmath$\Omega_c h^2 $} & \lcdmCMBextomch{} &  \neffCMBextomch{} & \yheCMBextomch{} & \yheneffCMBextomch{} \\

{\boldmath$100\theta_{MC} $} & \lcdmCMBexttheta{} &  \neffCMBexttheta{} & \yheCMBexttheta{} & \yheneffCMBexttheta{} \\

{\boldmath$\tau $} & \lcdmCMBexttau{} &  \neffCMBexttau{} & \yheCMBexttau{} & \yheneffCMBexttau{} \\

{\boldmath${\rm{ln}}(10^{10} A_s)$} & \lcdmCMBextlogA{} &  \neffCMBextlogA{} & \yheCMBextlogA{} & \yheneffCMBextlogA{} \\

{\boldmath$n_s $} & \lcdmCMBextns{} &  \neffCMBextns{} & \yheCMBextns{} & \yheneffCMBextns{} \\
\hline
{\boldmath \neff{}} & - &  \neffCMBextnnu{} & - &   \yheneffCMBextnnu{} \\
{\boldmath \yhe} & - & - &\yheCMBextyhe{} & \yheneffCMBextyhe{}\\
\hline
 {\boldmath $H_0$} (km s$^{-1}$ Mpc$^{-1}$) & \lcdmCMBextho{} &  \neffCMBextho{} & \yheCMBextho{} & \yheneffCMBextho{} \\
 
\hline\hline
\end{tabular} 
\label{table_cmbext}
%\vspace{8pt}
\tablecomments{
The median parameter values and 68\% confidence intervals with the CMBext dataset for the \lcdm{} model as well as the three extensions to the \lcdm{} model considered in this work. 
In the last row, we also show the constraints on a derived parameter, the Hubble constant, $H_0$, which has received attention recently due to tensions in the preferred value between different experiments. 
}
\end{center}
%\noindent Values of 6 parameters 68\% limits from the constraint of \lcdm with 3 combinations of datasets.
\end{table*}

\comment{
\begin{table*}
\begin{center}
\caption{Parameter constraints for \lcdm\ + $Y_{He}$ + $N_{eff}$}  \label{table_yheneff}
\hskip-4.0cm\begin{tabular} { l  c c c}

 & Planck  &CMB & CMBext\\
\hline
{\boldmath$\Omega_b h^2   $} & $0.02224\pm 0.00023         $ & $0.02221\pm 0.00022 $ & $0.02221\pm 0.00020 $\\

{\boldmath$\Omega_c h^2   $} & $0.1174^{+0.0041}_{-0.0047}$ & $0.1148^{+0.0039}_{-0.0046}$ & $0.1139^{+0.0035}_{-0.0043}$\\

{\boldmath$100\theta_{MC} $} & $1.0414\pm 0.0012 $  & $1.0424\pm 0.0012 $ & $1.0426\pm 0.0011$\\

{\boldmath$\tau           $} & $0.0536\pm 0.0079 $  & $0.0514\pm 0.0080 $ & $0.0506\pm 0.0075          $\\

{\boldmath$N_{eff}        $} & $2.86^{+0.27}_{-0.30} $  & $2.68^{+0.26}_{-0.30} $ & $2.63^{+0.23}_{-0.28} $\\

{\boldmath$Y_{He}         $} & $0.247^{+0.018}_{-0.016}   $ & $0.262\pm 0.016 $ & $0.264\pm 0.015            $\\

{\boldmath${\rm{ln}}(10^{10} A_s)$} & $3.037\pm 0.019 $  & $3.032\pm 0.019 $ & $3.027\pm 0.016            $\\

{\boldmath$n_s            $} & $0.9593\pm 0.0085 $  & $0.9590\pm 0.0084 $ & $0.9583^{+0.0074}_{-0.0084}$\\
\hline
\end{tabular} \par
\end{center}
\vspace{8pt}
\noindent Values of 6 parameters 68\% limits from the constraint of  \lcdm\ +$Y_{He}$+ $N_{Eff}$ with 3 combinations of datasets.
\end{table*}}

\section{Conclusions} \label{sec:final}
%0.5p

We have presented a measurement of the CMB \emode{} power spectrum on angular multipoles $500 \le \ell \le 3000$ from 670\,\sqdeg{} surveyed with the \pb{} instrument.  
\emode{} polarization is detected at high significance across the third through the seventh acoustic peaks of the \emode{} power spectrum. 
We find no evidence for significant systematic biases in the null suite data.  
The \pb{} \emode{} bandpowers provide an independent confirmation of the observed CMB \emode{} power spectrum at intermediate-to-small angular scales. 

We combine the \pb{} \emode{} bandpowers with other recent CMB measurements \citep{louis17, planck18-5, henning18} to explore the current state of CMB cosmological constraints. 
Adding the ground-based CMB bandpowers does not reduce the Hubble constant tension between the \planck{} inferred value and direct local measurements (4.3\,$\sigma$ vs 4.5\,$\sigma$). 
We find no significant preference in the data for any of the extensions considered: \neff, \yhe,  \neff+\yhe. 

For the \lcdm+\yhe{} model extension, adding the ground-based CMB power spectrum measurements brings the helium abundance towards the BBN expectation of 0.2467. 
With \planck-only, the data prefer $\yhe=\yhePlanckyhe$, shifting to $\yhe=\yheCMByhe{}$ when the other power spectrum measurements are added. 
As expected, the non-CMB-power-spectrum data does little for \yhe. 

We also look at varying the effective number of relativistic species, \neff. 
We find only minor improvements and shifts from adding data beyond the \planck{} bandpowers. 
For the combined CMBext dataset, the data favor $\neff = \neffCMBextnnu{}$ which is within $1\,\sigma$ of the expected value of 3.046. 

Finally, we allow both \yhe{} and \neff{} to vary to study the degeneracies between the two. 
Here, the full CMB dataset slightly pulls \yhe{} upwards and \neff{} downwards relative to the \planck{} constraints and expected values. 
We find for CMBext, $\neff = \yheneffCMBextnnu{}$ and $\yhe = \yheneffCMBextyhe{}$. 
However, the actual improvement in the quality of fit from adding these two parameters is small ($\Delta \chi^2 \simeq -1.2$) suggesting that these shifts are not significant.

While the \pb{} survey has finished, its successor, the Simons Array, had first light in 2019. 
The complete Simons Array will have three telescopes with a total of about 20 times more detectors than \pb{} and will survey a large fraction of the Southern sky \citep{2016JLTP..184..805S, MasayaSPIE}. The Simons Array will also extend the survey area to the North at the equator, facilitating studies of cross correlations with experiments at other wavelengths. 
The \emode{} power spectrum measurement from the Simons Array survey will dramatically improve upon current \emode{} constraints and enable new tests of cosmology. 

\acknowledgements

The \pb\ project is funded by the National Science Foundation under grants AST-0618398 
and AST-1212230.  The analysis presented here was also supported by Moore Foundation 
grant number 4633, the Simons Foundation grant number 034079, and the Templeton Foundation 
grant number 58724.
The James Ax Observatory operates in the Parque Astron\'omico Atacama in Northern Chile under 
the auspices of the Comisi\'on Nacional de Investigaci\'on Cient\'ifica y Tecnol\'ogica de Chile (CONICYT).

The Melbourne group acknowledges support from the University of Melbourne and an Australian Research Council's Future Fellowship (FT150100074).
AK acknowledges the support by JSPS Leading Initiative for Excellent Young Researchers (LEADER) and by the JSPS KAKENHI Grant Numbers JP16K21744 and 18H05539.
CB, NK, and DP acknowledge support from the ASI-COSMOS Network (cosmosnet.it) and from the INDARK INFN Initiative (web.infn.it/CSN4/IS/Linea5/InDark).
GF acknowledges the support of the European Research Council under the European Union's Seventh Framework Programme (FP/2007-2013) / ERC Grant Agreement No. [616170] and of the UK STFC grant ST/P000525/1. 
HN acknowledges JSPS KAKENHI grant JP26800125.
%
%JC is supported by the European Research Council under the European Union's Seventh Framework Programme (FP/2007-2013) / ERC Grant Agreement No. [616170].
%
MA acknowledges support from CONICYT UC Berkeley-Chile Seed Grant (CLAS fund) Number 77047, Fondecyt project 1130777 and 1171811, DFI postgraduate scholarship program and DFI Postgraduate Competitive Fund for Support in the Attendance to Scientific Events.
MD acknowledges funding from the Natural Sciences and Engineering Research Council of Canada and Canadian Institute for Advanced Research.
MH acknowledges the support from the JSPS KAKENHI Grant Numbers JP26220709 and JP15H05891.
NK acknowledges the support from JSPS Core-to-Core Program, A. Advanced Research Networks.
OT acknowledges the SPIRITS grant in the Kyoto University, and JSPS KAKENHI JP26105519.
ST was supported by Grant-in-Aid for JSPS Research Fellow JP14J01662 and JP18J02133.
YC acknowledges the support from the JSPS KAKENHI Grant Number 18K13558, 18H04347, 19H00674.
The APC group acknowledges the travel support from the Labex Univearths grant.
This work was supported by the World Premier International Research Center Initiative (WPI), MEXT, Japan.
This research used resources of the Central Computing System, owned and operated by the Computing Research Center at KEK. 
Support from the Ax Center for Experimental Cosmology at UC San Diego is gratefully acknowledged.
Work at LBNL is supported in part by the U.S. Department of Energy, Office of Science, Office of High Energy Physics, under contract No. DE-AC02-05CH11231.
%
%This work was supported by the Moore Foundation, the Templeton Foundation and the Simons Foundation.
%
This research used resources of the National Energy Research Scientific Computing Center, which is supported by the Office of Science of the U.S. Department of Energy under Contract No. DE-AC02-05CH11231.
%
%We acknowledge many useful conversations with Nathan Whitehorn. 
%
%We acknowledge the use of the \texttt{emcee} package \citep{emcee}.
%
%Some of the results in this paper have been derived using the HEALPix \citep{Gorski2005} package.

\appendix\label{appendix}
As mentioned in \S\ref{sec:systematics}, we used 19 null tests to search for hidden systematics. Fourteen of them were previously used in PB19 and listed here for the readers' convenience. \\

\begin{itemize}
\item ``First half versus second half'': the dataset is split into two equal-weight halves chronologically to probe for time-dependent changes in the instrument, such as drifting calibration 

\item \enquote{Middle versus rising and setting}: the three different CES types are split in middle range elevation scans versus rising plus setting scans to detect, for example, elevation-dependent miscalibration or residual ground synchronous signal.

\item \enquote{Left-going versus right-going subscans}: the dataset is split in half according to the direction of motion of the telescope to test for, for example, microphonic or magnetic pickup in the data.

\item \enquote{High gain versus low gain observations}: the dataset is split into observations with above and below average mean detector gain coefficients to search for problems with the gain calibration.

\item \enquote{High PWV versus low PWV}: the dataset is split by PWV as measured by the nearby \apex\ radiometer to check for loading or weather dependent effects.

\item \enquote{Mean temperature to polarization leakage by channel}: split the dataset into detectors that see small and large temperature leakage coefficients to test the subtraction and search for residual contamination.

\item \enquote{$2f$ amplitude by channel}, \enquote{$4f$ amplitude by channel}: split the data by HWP signal amplitude to check for problems removing the HWP structure or systematic contamination coupling into the data through these terms.

\item \enquote{$Q$ versus $U$ pixels}: each detector wafer is fabricated with two sets of polarization angles. We split the data into the two pixel types to check for problems in the device fabrication.

\item \enquote{Sun above or below the horizon}, \enquote{Moon above or below the horizon}: we split observations based on whether or not the sun or moon is up to check for residual sidelobe contamination.

\item \enquote{Top half versus bottom half}, \enquote{left half versus right half}: we split detectors by the boresight axis of the telescope to check for optical distortion and problems due to  far sidelobes.

\item \enquote{Top versus bottom bolometers}: with a continuous HWP each bolometer TOD independently measures $Q$ and $U$. We explicitly separate detector pairs to check for temperature aliasing or device mismatch.
\end{itemize}
The other 5 splits are:
\begin{itemize}
\item  \enquote{Mean temperature to polarization leakage by CES}: split the dataset into CESs that see small and large temperature leakage coefficients to test the subtraction and search for residual contamination.
\item \enquote{$2f$ amplitude by CES}, \enquote{$4f$ amplitude by CES}: split the data by HWP signal amplitude for CES to check for problems removing the HWP structure or systematic contamination coupling into the data through these terms.
\item \enquote{Low distance or high distance from Sun}: we split observations based on the distance to the Sun to check for residual sidelobe contamination.
\item \enquote{Random splits of bolometers}: we randomly split bolometers into 2 halves to check the noise model.
\end{itemize}

\bibliographystyle{aasjournal}
\bibliography{pb}
%\bibliography{../../Bibtex/pb}

%largepatch}

\end{document}